\documentclass{svjour3}                     
\smartqed  
\usepackage{graphicx}
\begin{document}

\title{
\vspace{-2.0cm}
\begin{flushright}
{\large \bf LFTC-18-2/23} 
\end{flushright}
Charm production in interactions of antiproton with proton and nuclei
at ${\bar P}ANDA$ energies 
}


\author{R.~Shyam \and K.~Tsushima  
}
\institute{
R.~Shyam \at
Saha Institute of Nuclear Physics, 1/AF Bidhan Nagar, Kolkata 700064, India\\
\email{radhey.shyam@saha.ac.in}
\and
K.~Tsushima \at
Laborat\'{o}rio de F\'{i}sica Te\'{o}rica e Computacional, Universidade
Cruzeiro do Sul, Rua Galv\~{a}o Bueno, 868, Liberdade 01506-000, S\~{a}o Paulo, 
SP, Brazil 
}

\date{Received: date / Accepted: date}

\maketitle

\begin{abstract}
We study the production of charmed baryons in the antiproton-proton and 
antiproton-nucleus interactions within a fully covariant model that is 
based on an effective Lagrangian approach. The baryon production 
proceeds via the $t$-channel $D^0$ and $D^{*0}$ meson-exchange diagrams. 
We have also explored the production of the charm-baryon hypernucleus 
$^{16}_{\Lambda_c^+}$O in the antiproton - $^{16}$O collisions. For 
antiproton beam momenta of interest to the ${\bar P}ANDA$ experiment, 
the 0$^\circ$ differential cross sections for the formation of 
$^{16}_{\Lambda_c^+}$O hypernuclear states with simple particle-hole 
configurations, have magnitudes in the range of a few $\mu$b/sr.  

\end{abstract}

\section{Introduction}
Several interesting and intriguing questions in hadron physics can be 
elucidated by experiments involving medium-energy antiproton (${\bar p}$) 
beams on fixed-targets. The future ${\bar P}ANDA$ ("antiproton annihilation 
at Darmstadt") experiment at the under-construction antiproton and ion 
research facility (FAIR) in Darmstadt, Germany, will perform such studies 
at the beam momenta $\leq$ 15 GeV/c~\cite{pan09}. This includes measurements 
of the charm-meson and charm-baryon production in the antiproton (${\bar p}$) 
collisions with protons and nuclei at the beam momenta $\leq$ 15 GeV/c. The 
accurate knowledge of the charm-meson ${\bar D} D$ (${\bar D}^0 D^0$ and 
$D^- D^+$) production cross sections is important because the charmonium 
states above the open charm threshold will generally be identified by means 
of their decays to ${\bar D} D$ channels~\cite{wie11}.  

Studies of production and spectroscopy of charm-baryons (e.g.~$\Lambda_c^+$) 
are similarly interesting. In contrast to the mesons, there can be more states 
of these systems as there are more possibilities of orbital excitations 
(baryon resonances) due the presence of three quarks. At higher ${\bar p}$ 
beam momenta at the ${\bar P}ANDA$ facility the yields of the channels with 
charm-baryons exceed those of the charm-meson channels by factors of 3-4, 
which is confirmed by calculations reported in Refs.
\cite{shy14,shy16b,hai10,hai14}.  

The $\Lambda_c^+ - N$ interaction has come in focus after discoveries of
many exotic hadrons [e.g, $X(3872)$, and $Z(4430)$] by the Belle
experiments~\cite{cho08}. Because performing scattering experiments in 
this channel is not feasible for the time being, a viable alternative to 
determine this interaction is provided by the studies of the $\Lambda_c^+$ 
hypernuclei that can be done by the ${\bar p}$ induced reactions on nuclei 
at the ${\bar P}ANDA$ facility. In the past the studes of the $\Lambda$-  
and $\Xi$-hypernuclear states have provided important information about the 
$\Lambda - N$ and $\Xi - N$ interactions, respectively \cite{gal16,shy12}. 
The existence of the $\Lambda_c^+$ hypernuclei was predicted already in 
1975~\cite{tya75}. More recently, systematic studies have been reported of 
the $\Lambda_c^+$ hypernuclei with mass numbers ranging between 17 to 209 
within the quark-meson coupling (QMC) model (see, e.g.~\cite{tsu04}). 

\begin{figure}[t]
\begin{center}
\includegraphics[width=0.55\textwidth]{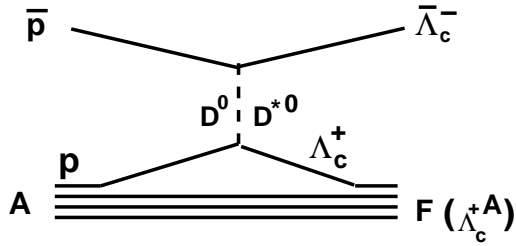}
\caption{Graphical representation of the model used to describe the
charmed baryon and charm-baryon hypernuclear production in ${\bar p} p$  
and ${\bar p} A$ collisions. In case of the ${\bar p} p$ collision the 
$\Lambda_c^+$ is a free particle in the final state, while in case of the 
${\bar p} A $ collision it gets captured to one of the nuclear orbits 
leading to the hypernucleus  $F ( = {_{\Lambda_c^+}}A)$. $D^0$ and $D^{*0}$ 
in the intermediate line represent the exchanges of $D^0$ pseudoscalar and 
$D^{*0}$ vector mesons, respectively.}
\vspace{-2em}
\label{fig1}       
\end{center}
\end{figure}
In this contribution, we present some results of our investigations of the 
production of charmed baryons in the antiproton-proton interactions and of 
the charm-baryon hypernucleus $^{16}_{\Lambda_c^+}$O in the ${\bar p}$ - 
$^{16}$O collisions, within a fully covariant model that is based on an 
effective Lagrangian approach.  
%

\section{Results and Discussions}

\subsection{Production of charmed baryons in ${\bar p}p$ collisions}

We have calculated the cross sections of the ${\bar \Lambda}_c^- 
\Lambda_c^+$,  ${\bar \Lambda}_c^- \Sigma_c^+$, and ${\bar \Sigma}_c^- 
\Sigma_c^+$ production channels in ${\bar p} p$ collisions within a 
single-channel effective Lagrangian model~\cite{shy14,shy17b,shy17a}), 
where this reaction is described as a sum of the $t$-channel $D^0$ and 
$D^{*0}$ meson-exchange diagrams (see, Fig.~\ref{fig1}). The effective 
Lagrangians for the $D^0$ and $D^{*0}$ meson-exchange vertices were 
taken to be ${\cal L}_{D^0BN} = ig_{BD^0N} {\bar \psi}_B \gamma_5 
\psi_N \phi_{D^0} + H.c.,$ and ${\cal L}_{D^{*0}BN} = g_{D^{*0}BN} 
{\bar \psi}_B \gamma_\mu \psi_N \theta_{D^{*0}}^\mu + \frac{f_{D^{*0}BN}}
{4M} {\bar \psi}_B \sigma_{\mu \nu} \psi_N G_{D^{*0}}^{\mu \nu} + H.c.$,
respectively. In these expressions $\psi's$ represent the baryon fields. 
$\phi$ and $\theta$ depict the fields of  $D^0$ and $D^{*0}$ mesons, 
respectively. The values of the coupling constants $g_{ND^0B}$, 
$g_{ND^{*0}B}$, and $f_{ND^{*0}B}$ were 13.98, 5.64 and 18.37, 
respectively, for vertices involving the $\Lambda_c^+$ baryon and 2.69,
3.25 and -7.88, respectively, for $\Sigma_c^+$ baryon vertices.  

\begin{figure}[t]
\begin{center}
\vspace{1.0em}
\includegraphics[width=0.53\textwidth]{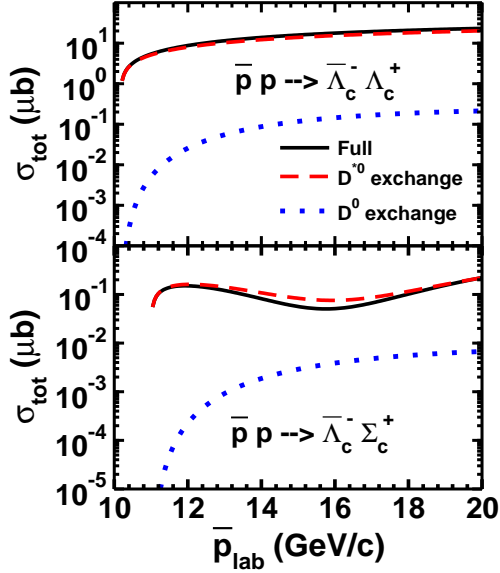}
\caption{Total cross section for ${\bar p}p \to {\bar \Lambda}_c^- \Lambda_c^+$ 
(upper pannel) and ${\bar p}p \to {\bar \Lambda}_c^- \Sigma_c^+$ (lower pannel) 
reactions as a function of the beam momentum.  
}
\vspace{-3em}
\label{fig2}       
\end{center}
\end{figure}

Fig.~\ref{fig2} displays the total cross sections of the reactions ${\bar p}p 
\to {\bar \Lambda}_c^- \Lambda_c^+$ [upper panel], and ${\bar p}p \to {\bar 
\Lambda}_c^- \Sigma_c^+$ [lower panel] as a function of ${\bar p}$ beam momenta
that vary in the range of threshold to 18 GeV/$c$,  which is of interest to the 
${\bar P}ANDA$ experiment. The threshold beam momenta for  ${\bar \Lambda}_c^- 
\Lambda_c^+$ and ${\bar \Lambda}_c^- \Sigma_c^+$ production channels are 10.162 
GeV/c and 10.99 GeV/c, respectively. For ${\bar p}_{lab}$ around 15 GeV/c, 
$\sigma_{tot}$ for the  ${\bar \Lambda}_c^- \Lambda_c^+$ channel is about one 
order of magnitude larger than that for the ${\bar \Lambda}_c^- \Sigma_c^+$ 
channel. The likely reasons for this difference are the smaller coupling 
constants at the $ND^{*} \Sigma_c^+$ vertices and the destructive interference 
between the $D^{*0}$ and $D^0$ exchange terms in case of the ${\bar \Lambda}_c^- 
\Sigma_c^+$ final state.

For both production channels, we note that the $D^{*0}$ exchange process 
dominates the cross sections. The  $D^0$ exchange contributions
are nearly two orders of magnitude smaller than those of the $D^{*0}$ exchange
in case of the ${\bar \Lambda}_c^- \Lambda_c^+$ final state and nearly an order
of magnitude for the ${\bar \Lambda}_c^- \Sigma_c^+$ final state in the region
of higher beam momenta. Interestingly, we notice in the lower panel that, even 
though for ${\bar p}_{lab}$ beyond 14 GeV/c the individual contributions of the 
$D^0$ exchange terms are at least one order of magnitude smaller, they still
influence the total cross sections significantly through the interference terms
that are destructive in this case.

\subsection{Production of charm-baryon hypernucleus $^{16}_{\Lambda_c^+}$O in
${\bar p}$ - $^{16}$O collisions}

We describe this reaction within an effective Lagrangian model presented above.
As discussed earlier, ${\bar \Lambda}_c^- \Lambda_c^+$ production takes place 
via $t$-channel exchanges of $D^0$ and $D^{*0}$ mesons in collisions of  
${\bar p}$ with one of the protons of the target nucleus in the initial state 
[see, Fig.\ref{fig1}]. The $\Lambda_c^+$ is captured into one of the orbits 
of the residual nucleus to make the hypernucleus, while ${\bar \Lambda}_c^-$ 
rescatters onto its mass shell. At the upper vertices of Fig.~\ref{fig1}, 
the amplitudes involve free-space spinors of the antiparticles, while at the 
lower vertices, they have spinors for the bound proton in the initial state and 
bound $\Lambda_c^+$ in the final state. These are the  solutions of the Dirac 
equation for a bound state problem in the presence of external potential fields. 
They are calculated within the QMC model. In this model~\cite{gui88}, quarks 
within the non-overlapping nucleon bags (modeled using the MIT bag), interact 
self-consistently with the isoscalar-scalar ($\sigma$) and isoscalar-vector 
($\omega$) mesons in the mean field approximation. The self-consistent response 
of bound quarks to the mean $\sigma$ field leads to a new saturation mechanism 
for nuclear matter. For a comprehensive review of this model and its 
applications, we refer to Ref.~\cite{sai07}.
\begin{figure}[t]
\begin{center}
\vspace{1em}
\includegraphics[width=0.53\textwidth]{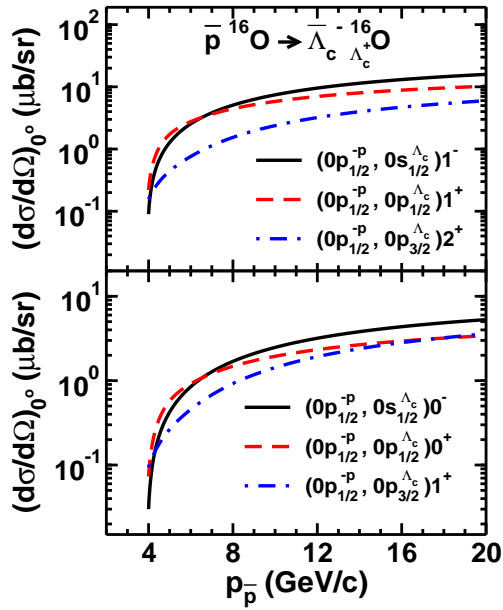}
\caption{
(Upper panel) Differential cross sections at $0^\circ$ of the ${\bar p}$ 
$^{16}$O $\to {\bar \Lambda}_c^- {^{16}_{\Lambda_c^+}}$O reaction leading to the
$^{16}_{\Lambda_c^+}$O hypernuclear states of larger $J$ value of each
particle-hole configuration as indicated. (Lowe panel) The same as in (a) but for
states of lower $J$ value of each configuration as indicated. In the
legends $\Lambda_c$ corresponds to $\Lambda_c^+$.
}
\vspace{-3em}
\label{fig3}       
\end{center}
\end{figure}

The QMC model predicts three bound states for the charm-baryon hypernucleus 
$^{16}_{\Lambda_c^+}$O. The predicted quantum numbers and binding energies of 
these states are: [${^{16}_{\Lambda_c^+}}$O($\Lambda_c^+$ 0$p_{1/2}$, BE = 7.17 
MeV)], [${^{16}_{\Lambda_c^+}}$O($\Lambda_c^+$ 0$p_{3/2}$, BE = 7.20 MeV)], 
and [${^{16}_{\Lambda_c^+}}$O($\Lambda_c^+$ 0$s_{1/2}$, BE = 12.78 MeV)]. We 
assume the initial bound proton state to have quantum numbers of the outermost 
$0p_{1/2}$ proton orbit of the target nucleus. The predicted binding energy of 
this state within the QMC model is 11.87 MeV.

In Fig.~\ref{fig3}, we show the $0^\circ$ differential cross sections
$[(d\sigma/d\Omega)_0]$ for the reaction
${\bar p}$ $^{16}$O $\to {\bar \Lambda}_c^-\, {^{16}_{\Lambda_c^+}}$O
obtained by using the proton-hole and $\Lambda_c^+$ bound state spinors
calculated within the QMC model. Cross sections are shown for ${\bar p}$
beam momenta in the range of threshold to 20 GeV/c. The charm-baryon 
hypernuclear states populated are $1^-$ and $0^-$, $1^+$ and $0^+$, and 
$2^+$ and $1^+$ corresponding to the particle-hole configurations 
($0p_{1/2}^{-p},0s_{1/2}^{\Lambda_c^+}$),
($0p_{1/2}^{-p},0p_{1/2}^{\Lambda_c^+}$), and
($0p_{1/2}^{-p},0p_{3/2}^{\Lambda_c^+}$), respectively. Cross sections
to the higher $J$ state of each configuration are shown in the upper panel
while those to lower $J$ in the lower panel.  We see that for each
particle-hole configuration, the state with higher $J$ has larger cross
section. For ${\bar p}$ beam momenta of interest to the ${\bar P}ANDA$ 
experiment (between 8 - 15 GeV/c), the magnitudes of $0^\circ$ differential 
cross sections vary between 1.5 - 3.8 $\mu$b/sr, and 5.0 - 11.0 $\mu$b/sr for 
states $0^-$ and $1^-$, respectively, of the configuration ($0p_{1/2}^{-p},
0s_{1/2}^{\Lambda_c^+}$). On the other hand, for states $1^+$ and $2^+$ of 
the configuration ($0p_{1/2}^{-p},0p_{3/2}^{\Lambda_c^+}$), it varies 
between 0.9 - 2.8 $\mu$b/sr, and 1.6 - 6.0 $\mu$b/sr, respectively. These 
are relatively substantial values. 

\section{Conclusion}

We investigated the production of charmed baryons, ${\bar \Lambda}_c^- \Lambda_c^+$, 
${\bar \Lambda}_c^- \Sigma_c^+$, in the ${\bar p}p$ collisions within an effective 
Lagrangian model that involves the meson-baryon degrees of freedom. The production 
mechanism is described by the $t$-channel $D^0$ and $D^{*0}$ meson-exchange diagrams,
while largely phenomenological initial- and final-state interactions have been used 
to account for the distortion effects. In the range of beam momenta of interest to 
the ${\bar P}ANDA$ experiment, the total cross sections for the  ${\bar \Lambda}_c^- 
\Lambda_c^ + $ production channel are about one order magnitude larger than those 
of the ${\bar \Lambda}_c^- \Sigma_c^+$ channel. The reasons for this is  large 
destructive interference between the vector and tensor parts of the $D^{*0}$ 
meson-exchange term and relatively smaller coupling constants of the 
$ND^{*0}\Sigma_c^+$ vertices.   

We have also studied the production of charm-baryon hypernucleus 
$^{16}_{\Lambda_c^+}$O in ${\bar p}$ - $^{16}$O collisions within a similar model.
At beam momenta of interest to the ${\bar P}ANDA$ experiment, the $0^\circ$ 
differential cross section for the ${\bar p}$ $^{16}$O$\to {\bar \Lambda}_c^-  
{^{16}_{\Lambda_c^+}}$O reaction varies between 0.9 $\mu$b/sr to 11 $\mu$b/sr depending 
on the final $\Lambda_c^+$ state excited in the reaction. This together with the low 
threshold beam momentum (3.953 GeV/c) for the production of the $^{16}_{\Lambda_c^+}$O 
hypernuclear states in the $\bar p$ - $^{16}$O reaction, could make it possible to perform 
such experiments at the ${\bar P}ANDA$ facility even in the beginning stages of the FAIR.


\begin{acknowledgements}
This work has been supported by the Science and Engineering Research 
Board (SERB), Department of Science and Technology, Government of 
India under Grant no. SB/S2/HEP-024/2013, and by FAPESP, Brazil, under 
Grants, no.~2016/04191-3, and no.~2015/17234-0, and CNPq, Brazil under 
Grants no.~400826/2014-3 and no.~308088/2015-8.
\end{acknowledgements}

\end{document}